\documentclass[amsmath,amssymb,aps,prb,reprint,superscriptaddress,nobibnotes]{revtex4-2}

\usepackage{amsmath}
\usepackage{amsfonts}
\usepackage{amssymb}
\usepackage{graphicx}
\usepackage{upgreek}
\usepackage[dvipsnames]{xcolor}
\usepackage{braket}
\usepackage{verbatim}

\makeatletter 
\renewcommand\@biblabel[1]{#1.} 
\makeatother

\definecolor{darkred}{rgb}{0.5, 0, 0}
\definecolor{darkgreen}{rgb}{0, 0.5, 0}
\definecolor{darkblue}{rgb}{0.1, 0.1, 0.7}

\usepackage[bookmarks=true,
            bookmarksnumbered=false,
            bookmarksopen=true,
            breaklinks=true,
            pdfborder={0 0 0},
            final=true,
            backref=none,
            colorlinks=true,
            linkcolor=darkblue,
            menucolor=darkblue,
            citecolor=darkblue,
            urlcolor=darkblue]{hyperref}

\newcommand{\micro}{${\upmu}$}

\newcommand{\upsub}[1]{_{\mathrm{#1}}}
\newcommand{\um}{$\,$\micro m}
\newcommand{\uev}{$\,$\micro eV}

\newcommand{\nlu}{\uev\um$^{2}$}

\begin{document}

\title{All-optical nonlinear phase modulation in open semiconductor microcavities}

\author{Fedor A. Benimetskiy}
\email[]{f.benimetskiy@sheffield.ac.uk}
\affiliation{University of Sheffield, S3 7RH, Sheffield, UK}
\author{Paul M. Walker}
\email[]{p.m.walker@sheffield.ac.uk}
\affiliation{University of Sheffield, S3 7RH, Sheffield, UK}
\author{Anthony Ellul}
\affiliation{University of Sheffield, S3 7RH, Sheffield, UK}
\author{Oleksandr Kyriienko}
\affiliation{University of Sheffield, S3 7RH, Sheffield, UK}
\author{Aristide Lemaître}
\affiliation{Université Paris-Saclay, CNRS, Centre de Nanosciences et de Nanotechnologies (C2N), 91120 Palaiseau, France}
\author{Martina Morassi}
\affiliation{Université Paris-Saclay, CNRS, Centre de Nanosciences et de Nanotechnologies (C2N), 91120 Palaiseau, France}

\author{Tommi Isoniemi}
\affiliation{University of Sheffield, S3 7RH, Sheffield, UK}
\author{Maurice S. Skolnick}
\affiliation{University of Sheffield, S3 7RH, Sheffield, UK}
\author{Jason M. Smith}
\affiliation{Department of Materials, University of Oxford, Oxford OX1 3PH, United Kingdom}
\author{Jacqueline Bloch}
\affiliation{Université Paris-Saclay, CNRS, Centre de Nanosciences et de Nanotechnologies (C2N), 91120 Palaiseau, France}
\author{Sylvain Ravets}
\affiliation{Université Paris-Saclay, CNRS, Centre de Nanosciences et de Nanotechnologies (C2N), 91120 Palaiseau, France}
\author{Dmitry N. Krizhanovskii}
\affiliation{University of Sheffield, S3 7RH, Sheffield, UK}

\begin{abstract}

We report an 80 times advancement in single particle light-by-light phase modulation using micrometer scale open-access optical cavities in the strong light-matter coupling regime. We achieve cross-phase modulation of up to 247$\pm$17 mrad per particle between laser beams attenuated to single-photon intensities, reaching the same order of magnitude achieved with nano-scale semiconductor quantum dots. We characterise the frequency, polarisation and time dependences of the nonlinear interaction underpinning the phase modulation and find that our results are in agreement with a theoretical model incorporating polariton-biexciton coupling. These advances extend the potential for quantum information processing and nonlinear quantum optics in strongly coupled light-matter systems, setting a new benchmark in the field without relying on atom-like emitters. Our findings suggest promising new avenues for scalable quantum optical technologies.

\end{abstract}

\maketitle



\section*{Introduction}
Strong effective interactions between photons, mediated by material nonlinear optical response, underpin solutions to a wide range of scientific and technological challenges from quantum computing and entanglement distribution ~\cite{Chang_2014, maring2024versatile, bao2023very, vajner2022quantum}, to the ultra-low power limit of classical all-optical processing~\cite{Zasedatelev_2021}, to novel approaches to quantum machine learning~\cite{Shen_2017}. This has inspired great effort to build photonic systems exhibiting cross-phase modulation (XPM) where one optical signal can strongly and deterministically modulate the phase of a separate optical signal at the single-photon level. Such single photon phase shifters underpin routes to applications~\cite{jankowski2024ultrafast} such as two-photon gates~\cite{brod2016passive,Schrinski2022,Heuck2020,Tiarks_2018,Scala_2024} for quantum information processing, photon number detectors and sorters~\cite{yang2022deterministic}, non-Gaussian state generation for metrology~\cite{yanagimoto2022onset, yanagimoto2020engineering, de_las_Heras_2025}, programmable quantum circuits~\cite{nielsen2024}, and quantum optical neural networks~\cite{Ewaniuk2023}.

Single photon phase shifts up to $\pi$ have been observed using single~\cite{turchette1995measurement,hacker2016photon} or ensembles of natural atoms~\cite{Tiarks2016,Sagona-Stophel2020}. For simplicity and scalability it is desirable to incorporate the nonlinear elements on semiconductor chips. Recently, semiconductor artificial atoms (quantum dots, QDs)~\cite{Somaschi_2016,Tomm_2024} have allowed XPM phase shifts of $\sim$ 600 mrad~\cite{Staunstrup_2024} and have already been utilised for quantum simulation~\cite{nielsen2024}. Phase shifts of order $300$ mrad are already sufficient for implementing quantum optical neural networks~\cite{Ewaniuk2023}. Spin-photon interfaces allow $\pi$ optical phase shifts conditional on the electronic state of a QD~\cite{Mehdi2024}, but this matter-photon interaction is a qualitatively different regime from the effective photon-photon interaction of XPM. Challenges remain, moreover, with scaling up to circuits containing mesoscopic numbers of dots with the same operating frequency.

An intrinsically more scalable approach is to couple light to two-dimensional excitons confined in quantum wells (QW), which have homogeneous energies on large (millimeter) spatial length scales. In the strong light-matter coupling regime this results in formation of part-photon part-exciton superposition states called exciton-polaritons. Confinement in all spatial dimensions can then be applied entirely through the photon degree of freedom using well established photonic fabrications techniques. Interactions between the excitonic components of the polaritons are at least 1000 times stronger than semiconductor nonlinearities in the weak light-matter coupling regime and several works have used them to manipulate photon number statistics within single optical modes~\cite{delteil2019towards,munoz2019emergence, ordan2024electrically, schnuriger2026quantum}. Additionally, interactions between polaritons with the same or opposite spin have different strengths, being characterised by interaction constants $\alpha_1$ and $\alpha_2$ respectively. When the polariton frequency is close to half the QW biexciton transition $\alpha_2$ can be strongly enhanced~\cite{Takemura2014,takemura2017spin}. Coupling to biexcitons has been suggested to explain the strength of interaction induced polariton quantum correlations~\cite{Bleu2021,scarpelli2024probing, schnuriger2026quantum}.

A further advantage is that the nonlinear bandwidth, related to the light-matter coupling rate or vacuum Rabi splitting $\Omega\upsub{Rabi}$, exceeds several meV allowing ultrafast (picosecond) operation. The micrometer size of the optical modes makes coupling light in and out relatively simple. The tradeoff for these advantages is the smaller interaction strength compared to QDs. In a previous demonstration in solid microcavities~\cite{kuriakose2022few} we showed XPM of 3 mrad per particle, limited by large cavity size, low excitonic content of the polaritons, polariton frequencies detuned away from the biexciton resonance, and broadening of polariton linewidth due to etching through the QW. Despite these limitations the result was already an order of magnitude higher than any other systems not relying on atom-like emitters.

In this work, we experimentally study XPM in a tunable open-access polariton microcavity~\cite{dufferwiel2014strong} with significantly tighter spatial confinement (beam waist 1\um{}) and narrower excitonic resonances allowing smaller polariton linewidths (60\uev{}) at the higher excitonic content (55\%) where interactions are maximised. By these means we significantly advance the state of the art single polariton XPM by a factor of 80, up to 247$\pm$17 mrad. This is already in the regime useful for quantum optical neural networks~\cite{Ewaniuk2023}, with the homogeneity of QW polaritons potentially allowing many nonlinear nodes. We furthermore show modulation of a probe by 500 mrad (50\%) using fewer than 10 control particles, corresponding to a peak intra-cavity energy $<2$ aJ. We use our highly flexible platform to study time-delay, frequency detuning, polarisation and power dependences and address the microscopic physics underlying these giant interactions.
We find that there is negligible contribution from processes with a long lifetime ($>\sim$ 300 ps) such as, for example, slow excitonic reservoirs~\cite{Stepanov2019,Menard2014}.
Furthermore, we find that the strength of the nonlinear response is significantly enhanced and has a stronger frequency dependence than expected for interactions mediated by exciton-exciton repulsion alone. We achieve good agreement by including enhancement of interactions via polariton-biexciton coupling in our model.

\begin{figure}
\includegraphics{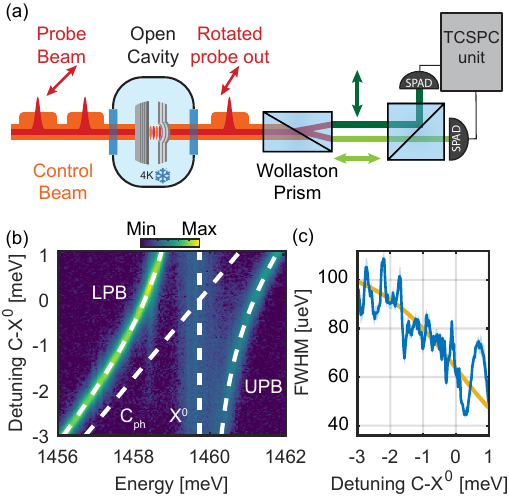}
\caption{\label{fig1} (a) Simplified schematic of the experimental setup. (b) White light transmission spectra as a function of the detuning between bare cavity mode ($C_{ph}$) and exciton ($X^0$) plotted in logarithmic colour scale. LPB (UPB) indicates the lower (upper) polariton branch. White dashed lines show a fit by a coupled oscillator model (further details in Supplementary Discussion 2B). (c) Polariton linewidth extracted from the white light transmission measurements, as a function of the detunings of the cavity mode from the exciton resonance (blue curve) and fit using the exciton and photon fractions from the coupled oscillator model (orange curve).}
\end{figure}

\begin{figure*}
\includegraphics{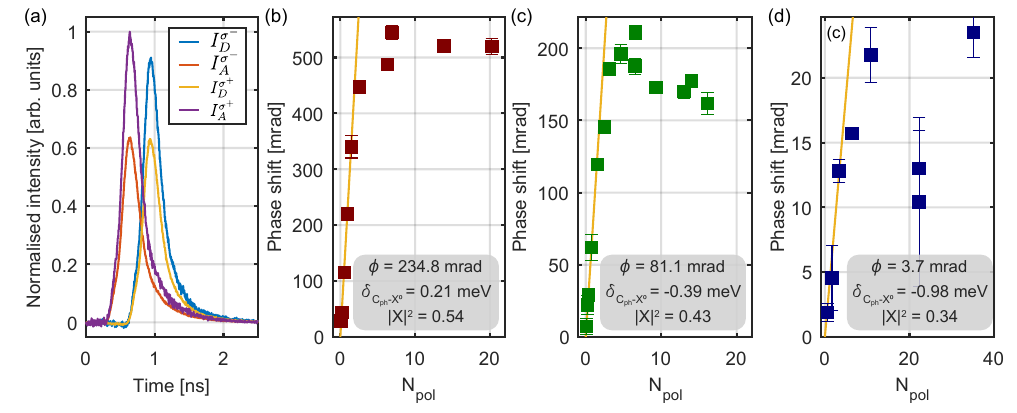}
\caption{\label{fig2}(a) Example time-correlated single photon counting (TCSPC) traces recorded during a phase-shift measurement. The control-beam background has been subtracted. A and D denote the SPAD channels measuring the diagonal and anti-diagonal polarisation components. $\sigma^{+}$ and $\sigma^{-}$ indicate the control-beam polarisation state. (b–d) Measured phase shift as a function of the control beam mean polariton number for different cavity–exciton detunings.}
\end{figure*}

\section*{Results}

Our measurements are performed in a polaritonic open-access microcavity~\cite{dufferwiel2014strong}  (see Methods~\ref{methods:sample}). Figure~\ref{fig1}b shows a white light transmission spectrum as the separation between the mirrors is varied, revealing clear anti-crossing between lower and upper polariton branches (see Methods~\ref{methods:sample}). The vacuum Rabi splitting $\Omega_{\mathrm{Rabi}} = 2.87$~meV between the two branches characterizes the light-matter coupling strength. The excitonic (photonic) content of the polaritons $\left|X\right|^2$ ($\left|C\right|^2 = 1-\left|X\right|^2$) increases (decreases) as the lower polariton energy approaches the exciton resonance.

Figure~\ref{fig1}c presents the lower polariton linewidth $\gamma$, which reduces significantly from $113 \pm 9$\uev{} in the photonic regime to $\sim 60$ \uev{} at high exciton fraction $\left|X\right|^2=$ 54\%. This indicates that for the detunings we study the linewidth is dominated by photonic losses with negligible contribution from the exciton. The ability to reach these narrow linewidths for for highly excitonic (and therefore nonlinear) polaritons is one of the key factors in achieving high single photon phase shifts. The cavity eigenstates in the absence of nonlinearity are linearly polarised along directions labelled $H$ and $V$ with an energy of splitting 0.86 times the linewidth $\gamma$ due to birefringence.

\begin{figure}
\includegraphics{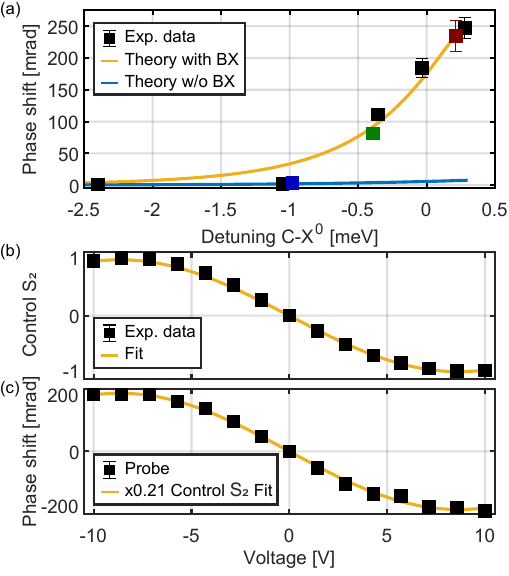}
\caption{\label{fig3}  (a) Phase shift per polariton as a function of cavity–exciton detuning. The center of the polariton-biexciton resonance occurs for cavity exciton detuning of 0.92 meV. For the theory w/o BX curve, $g_{\mathrm{eff}} = 10$~\nlu{}, as in Ref.~\cite{kuriakose2022few}.(b) $S_{2}$ Stokes parameter of the transmitted control beam vs. EOM voltage $V$. The orange line is a sinusoidal fit. (c) Probe phase shift corresponding to the control polarisation in (b) for 0.8 control polaritons in the cavity. The zero phase shift on the vertical axis corresponds to the probe polarisation state with no control beam present. The orange line is the same curve as in panel (b) but scaled by 0.21, indicating that the probe phase shift is proportional to the control $S_2$ Stokes parameter.}
\end{figure}

The phase rotation measurement is illustrated in Figure~\ref{fig1}a. 
Control and probe pulses of duration 1.3 ns and 0.3 ns respectively were generated from independent lasers and used to excite the fundamental cavity mode (see Methods~\ref{methods:pulses}). The peak number of control polaritons in the cavity was varied between 1 and 40. The probe pulses were $H$ polarised. The polarisation of the control beam incident on the cavity was modulated between left and right circular polarisations ($\sigma^+$ and $\sigma^-$) using an electro-optical modulator (EOM). Filtering of the incident control field by the birefringent-split cavity modes meant that the corresponding transmitted control polarisations were close to diagonal ($D$) and anti-diagonal ($A$) respectively (see Fig.~\ref{fig3}(b) and Supplementary Discussion 5). As usual for polariton cavities the transmitted polarisation is the same as the polarisation state inside the cavity. The polaritonic interactions in the cavity cause the control pulses to rotate the direction of the probe linear polarisation state away from $H$, which mathematically corresponds to a phase shift $\phi$ between circular polarisation components (see Supplementary Discussion 5). Changes in probe polarisation were measured by recording the intensities $I_{D,A}^{\sigma^+,\sigma^-}$ of diagonal  and anti-diagonal probe polarisation components for the $\sigma^+$ and $\sigma^-$ incident control polarisations (see Methods~\ref{methods:detection}). An example of the probe pulse temporal profiles, obtained after subtracting the control-beam background in post-processing (see Methods~\ref{methods:detection}), is shown in Fig.~\ref{fig2}(a).

One of the main advantages of using an open cavity system is that the polariton energy can be varied via the cavity-exciton detuning $\Delta$ allowing us to measure the frequency dependence of the nonlinearity. Figures~\ref{fig2}~b-d display the power dependencies of the phase shift for three detunings $\Delta$. All presented results demonstrate similar behaviour: the phase shift increases linearly with the number of polaritons in the cavity and saturates after reaching 5-10 polaritons. The value of the phase shift per polariton was extracted as the slope of the linear fit to the linear part of the graph for each detuning. The values are summarised in Fig.~\ref{fig3}a for 8 measurements over a wide range of detunings. The highest phase shift observed was 247$\pm$17 mrad per polariton for an excitonic fraction of 55\%. These results indicate a 80x times enhancement in phase shift per polariton compared to that reported in a monolithic air-post microcavity~\cite{kuriakose2022few}. Moreover, the phase shift reaches 500 mrad with $<10$ control polaritons in the cavity (see Fig.~\ref{fig2}b), corresponding to an intra-cavity energy of $<2$ aJ. This absolute phase shift is $>30$ times higher than that demonstrated in polariton micropillars, where the phase shift was limited to 15 mrad by saturation~\cite{kuriakose2022few}.

We have also examined the dependence of the phase shift on the control beam polarisation. 
For this part the control beam was scanned through 15 polarisation states and we recorded both the phase shift and the $S_2$ Stokes parameter of the transmitted control beam. We investigated 4 detunings in the range from –1.05 meV to 0.28 meV for $\leq$ 1.5 control polaritons. Fig.~\ref{fig3}b shows the control $S_2$ parameter as a function of the voltage $V$ applied to the EOM for a typical case (detuning of 0.28 meV and cavity occupancy of 0.8 control polaritons). We find that at zero voltage both the incident and transmitted control are $H$ linearly polarised ($S_{2,inc}=S_{2,tr}=0$). For the highest voltages $\left|S_{2,tr}\right|>0.9$ corresponding to an elliptically polarised field with a high linear polarisation degree. We note that even with $S_{2,tr}=0.9$ the value of $S_{3,tr}$ can be as high as $\sqrt{1-0.9^2}=0.44$. The solid curve is a sinusoidal fit.

Our model of the effect of the control on the probe predicts that the phase shift should be proportional to the control $S_{2,tr}$.
Figure~\ref{fig3}c shows the phase shifts corresponding the control polarisations in ~\ref{fig3}b. The solid curve is the same as that in ~\ref{fig3}b apart from a constant scaling factor of 0.21. This indicates that the phase shift follows the control polarisation as expected. This was also the case for the other detunings studied.

We now consider the dependence of the phase shift on the cavity-exciton detuning (Fig.~\ref{fig3}a). For detunings below -0.5 meV the phase shift reduces markedly, reaching a value of only 3.8 mrad at a detuning of -1 meV and remaining low for more negative detunings. The solid blue curve shows the expected phase shift mediated by the exciton-exciton interaction mechanism in our previous work~\cite{kuriakose2022few}. It significantly underestimates the phase shift. The frequency dependence comes from the $|X|^4$ dependence of the interaction strength on detuning~\cite{kuriakose2022few} and the frequency dependent polariton linewidth. This underestimates the observed strength of the frequency dependence, predicting a ratio of 2 between the highest phase shift and that at detuning -0.39 meV compared to the experimental ratio of 3. Explaining these features requires some additional frequency dependence of the polariton interactions. We will consider possible origins of this effect in the Discussion section below.

\begin{figure}
\includegraphics{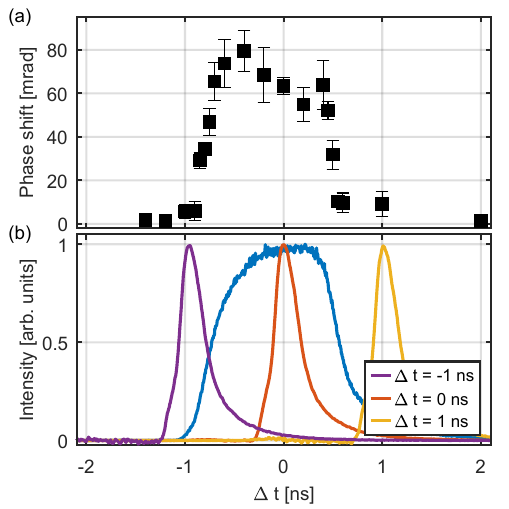}
\caption{\label{fig4} (a) Phase shift dependence on time delay between probe and control pulses. (b) Control pulse (blue) and probe pulse for three different delays. Note that the pulse shapes are convolved with the timing jitter of the SPADs.}
\end{figure}

Finally, we measured the response time of the nonlinearity underpinning the phase shift. Fig.~\ref{fig4}b shows the control temporal envelope along with the probe for three different relative delays. The phase shift vs. delay is shown in Fig.~\ref{fig4}a. For an instantaneous nonlinearity it is expected to be the same shape as the control pulse. A number of studies have demonstrated that even for resonant excitation polariton nonlinearities can be affected by a reservoir of dark exciton states whose occupation builds up and decays on timescales of a few hundred picoseconds~\cite{Vishnevsky2012,Sekretenko2013,Walker2017,Stepanov2019,del_Valle_Inclan_Redondo_2024}. For nonlinearities dominated by long-timescale effects the phase shift vs. delay is expected to exhibit an exponential rise (fall) on the leading (trailing) edge corresponding to build-up (decay) of the reservoir. In our case, the rise and fall times (between 10\% and 90\% of the peak of the phase shift) are 290 ps and 210 ps, respectively. These agree within 15 ps with the rise and fall times of the control pulse intensity measured independently using a streak camera.

This indicates that the nonlinearity causing the phase shift does not originate from reservoirs with timescales long compared to $\sim$300~ps. This is consistent with the negligible influence from reservoir effects recently observed in a different context but using cavities with very similar QWs~\cite{Frerot2023}. However, we cannot exclude contributions from reservoirs with fast population and decay times~\cite{Sekretenko2013,Walker2017}.

\section*{Discussion}
We have seen that the interactions leading to the phase shift rapidly decrease below detuning -0.5 meV, which cannot be accounted for by the varying excitonic fraction of the polaritons. The interactions also do not arise from particle accumulation in reservoir states with excitation/decay timescales $>300$ ps. We now consider a possible explanation for the strong frequency dependence. A series of studies~\cite{vladimirova2010polariton,takemura2017spin} has demonstrated that in the presence of biexcitons the ratio ${\alpha_2}/{\alpha_1}$ begins to depend on the cavity detuning with ${\alpha_2}$ being dramatically enhanced for polariton frequencies close to the biexciton resonance. Consequently, in experiments focusing on correlation spectroscopy~\cite{schnuriger2026quantum,delteil2019towards,munoz2019emergence,scarpelli2024probing} and XPM~\cite{kuriakose2022few} of polaritons observable effects can be strongly modified~\cite{Bleu2021}. This provides a possible source of nonlinearity with resonant behaviour and fast timescale. It should be noted that biexcitons cannot be generated using co-circularly polarised beams. However, in our experiment, the transmitted probe and control beams are linearly and elliptically polarised respectively, ensuring the presence of cross-polarised polaritons.
The interaction enhancement is expected when the polariton energy is within about a biexciton linewidth $\gamma\upsub{BX}$ of half the biexciton energy $E\upsub{BX}$. For $E\upsub{BX}=2.2$ meV and $\gamma\upsub{BX}=0.7$ meV respectively~\cite{takemura2017spin,scarpelli2024probing} the resonance occurs for $E\upsub{LP}-E_{X} = -1.1 \pm 0.7$ meV. With our $\Omega\upsub{Rabi}$ this range corresponds to cavity-exciton detunings $\Delta$ from -0.66 meV to +4.75 meV. We observed enhanced nonlinearity exactly in this region below the biexciton resonance peak. We could not measure higher detunings due to the onset of high absorption close to the exciton, which is often observed in polariton systems~\cite{delteil2019towards,munoz2019emergence,scarpelli2024probing}. 

Let us now motivate the results from the theoretical perspective.
In our system the Kerr interaction depends sensitively on the detuning $\Delta$ and the excitonic fraction $|X|^2$. The per-particle interaction energy between co-circularly polarized polaritons is given by $U_1 = \alpha_1/A\upsub{eff}$ where the interaction constant $\alpha_1$ is defined as 

\begin{equation}
\label{eq:alpha1}
\alpha_1(\Delta)=|X(\Delta)|^4\,g_{\mathrm{XX}}^{\parallel},
\end{equation}

Here $g_{\mathrm{XX}}^{\parallel}$ is the bare exciton–exciton interaction constant for parallel spins.
$A\upsub{eff}$ is the effective transverse mode area for nonlinear interactions as commonly used in nonlinear fiber optics~\cite{Agrawal2013} and in other works on polariton few-particle interactions~\cite{delteil2019towards,kuriakose2022few}. The cross-circularly polarized interaction constant, $\alpha_2$, can be derived in a similar way, but due to opposite spin configurations its exchange contributions are small and typically $|\alpha_2| \ll |\alpha_1|$ for GaAs samples at strong coupling.
However, in single QW samples where there is low inhomogeneous broadening narrow linewidth polaritons can be obtained close to the exciton where they are in resonance with the biexciton transition, allowing measurable polariton-biexciton coupling effects. These leads to an additional contribution to the interaction originating from a conversion process of two cross-circularly polarised excitons into one biexciton~\cite{Takemura2014}. The interaction energy between cross-circularly polarised polaritons then reads $U_2 = \alpha_2/A\upsub{eff}$ with

\begin{equation}
\label{eq:alpha2}
\alpha_2(\Delta)=|X(\Delta)|^4\left[g_{\mathrm{XX}}^{\perp}+\frac{g_{\mathrm{BX}}^2}{2E_{\mathrm{LP}}(\Delta)-E_{\mathrm{BX}}+i\gamma_{\mathrm{BX}}/2}\right],
\end{equation}

with $g_{\mathrm{XX}}^{\perp}$ being the bare interaction for antiparallel spins and $g_{\mathrm{BX}}$ the coupling strength between two excitons and the biexciton state.

The resonant behaviour of the second term as $2E_{\mathrm{LP}}(\Delta)$ approaches $E_{\mathrm{BX}}$ causes $\alpha_2$ to change much more rapidly than the $|X(\Delta)|^4$ scaling alone.

The actual phase shift imposed by the control depends not just on the strength of the nonlinearity but also on the polariton linewidth and birefringent splitting, the laser detuning from the polariton states, and (as seen above) the control polarisation state. In Supplementary Discussion 5 we derive an analytic expression which can be used to fit the phase shift vs. $\Delta$. The result of this fitting is shown as the solid curve in Fig.~\ref{fig3}a. The fixed parameters in the model are detailed in Methods~\ref{methods:model_params}. We allowed $\gamma\upsub{BX}$ and $g\upsub{BX}$ to be free parameters in the fit and obtained the best fit with values $\gamma\upsub{BX}=0.64$ meV and $g\upsub{BX}/\sqrt{A\upsub{eff}}=0.182$ meV.
These are in good agreement with the values of $\gamma\upsub{BX}=0.68$ meV and $g\upsub{BX}/\sqrt{A\upsub{eff}}=0.1$ meV from Ref.~\cite{scarpelli2024probing} suggesting that biexciton coupling is a reasonable candidate to explain our observations.

\begin{table}
\centering
\begin{tabular}{ |c|c|c|c|c|c| }
 \hline
 Ref. & $\left|U\right|$ (\uev{}) & $\left|X\right|^{2}$ & $A\upsub{eff}$ \um{}$^{2}$ & $g\upsub{eff}\left(\Delta\right)$ (\nlu{}) \\
 \hline
 This Work & 4.7 & 0.49 & 7.1 & 134 \\
 \hline
 \cite{munoz2019emergence} & 4.6 & 0.51 & 4.3 & 76 \\
 \hline
 \cite{scarpelli2024probing} & 2.7 & 0.42 & 4.3$^{*}$ & 66 \\
 \hline
 \cite{delteil2019towards} & 4.6 & 0.6 & 3.14 & 40 \\
 \hline
  \cite{schnuriger2026quantum} & 1.2 & 0.72 & 3.7 & 8 \\
 \hline
\end{tabular}
\caption{Comparison of nonlinear energies $U$ and effective interaction constants $g\upsub{eff}$ between tuneable polariton cavity systems in the few-particle limit. We have used $A\upsub{eff}=\pi w_{0}^2$ where $w_{0}$ is the beam waist (half width at $e^{-2}$). $^{*}$ Ref.~\cite{scarpelli2024probing} does not give a beam waist so we assume the same value as Ref.~\cite{munoz2019emergence} from the same group. In fact the mirror radius of curvature is $\sim1.4$x larger  so $g\upsub{eff}$ is likely to be larger.}
\label{tab:comparison}
\end{table}

Table~\ref{tab:comparison} compares representative polariton interaction energies $U$ measured in the available literature employing open-access microcavities, which are the most comparable systems to ours (see methods~\ref{methods:geff} for the definition of $U$). We also compare $g\upsub{eff}\left(\Delta\right) = \left|U\right|A\upsub{eff}/\left|X\right|^4$ which is the interaction energy with the dependencies on cavity mode area $A\upsub{eff}$ and exciton fraction scaled out. The values can still depend on frequency via processes such as biexciton coupling. There is significant variation among the literature values, although most show an enhancement compared to $g\upsub{eff}=g_{\mathrm{XX}}^{\parallel}/2\sim 5$\nlu{} expected for exciton-exciton interactions alone~\cite{Estrecho2019,kuriakose2022few}. Our value is $\sim2\times$ higher than those in Refs.~\cite{munoz2019emergence,scarpelli2024probing}.
Variation among works can be expected if the enhancement arises from biexciton coupling since the precise enhancement factor depends on the polariton-biexciton detuning at which measurements were made, as well as the details of the measurement itself. The works cited in Table~\ref{tab:comparison} indeed show a strong detuning dependence around the single values given in the table. The accessible detuning range is ultimately limited by QW disorder scattering for polariton frequencies close to the exciton~\cite{Diniz2011}, which depends on the details of the QW used. Furthermore, the value of $g_{BX}$ can also depend on the exciton density~\cite{vladimirova2010polariton,takemura2017spin} although we are not aware that this dependence has been observed experimentally. Finally, we note that although the biexciton coupling is a plausible candidate to explain the strong frequency dependent interaction we observe we cannot yet rule out other possible mechanisms. These could include enhancement of interactions by coupling of the polaritons to a reservoir of dark excitons with rapidly varying spectral density of states~\cite{Diniz2011} and timescale faster than $\sim 300$ ps~\cite{Sekretenko2013,Walker2017}, or mediation of coherent polariton-biexciton coupling by dark states~\cite{Fumero2025} (see supplementary discussion 6). Such mechanisms may produce sample-dependent nonlinearities with frequency dependences deviating from that expected for the simple polariton-biexciton coupling we analysed above.

\section*{Conclusion}
In conclusion, we have demonstrated light-by-light phase modulation at the few particle level, reaching up to 247$\pm$17 mrad per polariton, which is comparable to values reached using quantum dots and of order 1000x higher than in systems not using atom-like emitters. The single particle phase shifts are at the level useful for quantum enhanced processing devices such as quantum optical neural networks. With more control particles we reach absolute phase shifts of 500 mrad, surpassing previous achievements in polariton devices by $>30\times$. The significant enhancement of the phase shift compared to previous works is consistent with interaction enhancement by strong spatial confinement and the combination of narrow polariton linewidth and strong polariton-biexciton coupling at detunings close to the exciton, which can be simultaneously achieved in this sample due to the low inhomogeneous broadening of the QW. Further investigation is however needed to fully understand the details of the biexciton mechanism. This study marks a significant step forward, opening up possibilities for the development of quantum optical devices based on polariton interactions.

\section*{Methods}

\subsection{Sample and cavity properties}\label{methods:sample}
Our microcavity is formed between a dielectric concave distributed Bragg reflector (DBR) mirror with 21\um{} radius of curvature and a planar semiconductor sample. The semiconductor sample contains a 17 nm wide InGaAs quantum well (QW) in a GaAs spacer layer above a 33 layer-pair Al$_{0.1}$Ga$_{0.9}$As/AlAs DBR. The dielectric and semiconductor components are separated by an air gap, with their distance controlled by piezo positioner stages. Such hemispherical cavities support a lowest order mode with a tightly confined beam waist close to the quantum well position~\cite{dufferwiel2014strong}. Further details of the mirrors and cavity mode characterisation are given in Supplementary Discussion 1.

To minimise absorptive losses in the semiconductor material the whole cavity is kept at a temperature of 4 K by a helium exchange gas coupled to a bath of liquid helium. Details of the general optical setup are given in Supplementary Discussion 2A. We extracted the lower polariton dispersion and linewidth from white light transmission spectra at each detuning (blue curve in Fig.~\ref{fig1}c). The dispersion is fit with a coupled photon-exciton oscillator model. We then fit the linewidths with the model $\gamma = \left|C\right|^2\gamma_{C} + \left|X\right|^2\gamma_{X}$ where $\gamma$ and $\gamma_C$ and $\gamma_X$ are the polariton, photon and homogeneous exciton linewidths respectively and the exciton and photon fractions come from the coupled oscillator model. We obtain $\gamma_C = 113 \pm 9$\uev{} and  $\gamma_X = 14 \pm 4$\uev{} consistent with polariton linewidths dominated by photonic loss. This is consistent with previous measurements on samples with similar 17 nm wide InGaAs QWs~\cite{Frerot2023} and can be explained by the well-known phenomenon of motional narrowing~\cite{whittaker1996motional,savona1997microscopic,houdre1996vacuum}. The model accurately describes the linewidth of polariton with frequencies sufficiently below the exciton such that its inhomogeneous broadening does not play a significant role. This is the case for the detunings used in this work as can be seen from the good agreement between model and experiment in Fig.~\ref{fig1}(c).

For several cavity-exciton detunings between -7 meV and 0 meV we also confirmed the linewidth by scanning the cavity mode energy through a fixed single-frequency laser line and observing the Lorentzian shaped transmission vs. energy peaks. The measured linewidths agree with those from the white light transmission within the uncertainty. This measurement was also used to characterise the energy splitting and polarisation of the polariton eigenstates. We found a splitting of $0.858\gamma$ (splitting of $97\pm12$\uev{} in the purely photonic limit) between two linearly polarised eigenstates oriented in directions we will label $x$ and $y$. This splitting likely arises due to strain in the semiconductor sample~\cite{Tomm2021} or to small ellipticity of the concave mirror~\cite{Uphoff2015}. The ratio of splitting to linewidth was independent of detuning within measurement uncertainty which is likely because the linewidth is dominated by the photonic component so that both are approximately proportional to the photonic fraction of the polaritons. Further details of the dispersion, linewidth, and birefringent splitting measurment and the fitting are given in Supplementary Discussion 2B.

\subsection{Control and probe pulse generation}\label{methods:pulses}
We generate pump and probe pulses using electronically triggered high-speed fiber-based electro-optical modulators (EOMs) to carve single-frequency continuous wave lasers into pulses. This approach allowed precise control over the repetition rate, pump-probe delay, and pulse width. We used pulses with FWHM 300 ps (probe) and 1.3 ns (control). The rise/fall times of the pulses $<200$ ps are much longer than the cavity lifetime $\sim 10$ ps indicating that the spectral content of the pulses is significantly smaller than the linewidth or birefringent splitting.
The incident pulses were tuned in frequency to the center of the two birefringence-split polarisation states comprising the lowest lying transverse manifold. The control and probe beams had the same frequency (corresponding to the maximum transmission) but differing pulse temporal envelopes and were derived from separate lasers so that they had a random relative phase.

The pulses were attenuated to produce peak intra-cavity photon numbers in the range 1-40. The number of polaritons in the cavity $N\upsub{pol}$ was obtained from the transmitted power $P\upsub{out}$ using the formula

\begin{equation}\label{eq:num_pol}
    \hbar\omega N\upsub{pol} = P\upsub{out}\frac{t\upsub{rt}}{T|}\frac{1}{C|^2}
\end{equation}
where $|C|^2$ is the photon fraction of the polaritons, the transmission of the dielectric output mirror $T=1.4\times10^{-3}$ was measured using a laser and power meter, and the round trip time of photons in the cavity $t_{rt}=140$ fs was calculated using a transfer matrix method. Further details of the polariton number calibration are given in Supplementary Discussion 3.

The probe beam was linearly polarised with polarisation aligned to one of the birefringent split polarisation states (horizontally ($H$) polarised in the laboratory frame at the sample position). The polarisation of the control beam incident on the cavity could be tuned continuously between left and right circular, passing through $H$ linear polarisation, by means of a free-space EOM acting as a variable waveplate. After filtering by the birefringent split cavity states the transmitted control polarisation tunes from close to diagonal to close to anti-diagonal, passing through $H$. In polariton system the polarisation inside the cavity matches the transmitted one. For the results in Fig.~\ref{fig2} and Fig.~\ref{fig3}a the EOM was used to rapidly swap between left and right circular incident polarisation ($D$ and $A$ transmitted polarisation). For the results in Fig.~\ref{fig3}b the EOM was used to repeatedly scan the control polarisation through 15 points every 7.5 ms allowing a parameter sweep immune from gradual experimental drifts. Further details of the control and probe pulse generation are given in Supplementary Discussion 2C.

\subsection{Phase shift detection}\label{methods:detection}
After being transmitted through the cavity the output polarisation state is resolved in the diagonal (D) / anti-diagonal (A) basis (at the sample position) by a half wave plate and Wollaston prism polarizer allowing maximum sensitivity to changes in the probe polarisation direction. The transmitted photons were detected by single photon avalanche diodes (SPADs) and sorted into time bins using time-correlated single photon counting (TCSPC) electronics (PicoQuant HydraHarp) synchronised to the laser pulse triggering. This enabled us to record the pulse temporal envelopes with high signal to noise ratio and high temporal resolution and thus separate the pump and probe by temporal shape. Measurements were taken with the probe beam both on and off and the reference control pulse shape was subtracted to leave only the probe pulses. External marker channel inputs to the TCSPC electronics also allowed photon counts to be binned according to control beam polarisation state allowing high speed (kHz) repetitive polarization dependence sweeps. Further details of the detection setup are given in the second half of Supplementary discussion 2C.

Without the control beam present the $H$ polarised probe results in equal count rates on the SPADs measuring the $D$ and $A$ polarization components. The phase shift $\phi$ can be deduced from the change in the count rates and in particular by the diagonal polarisation degree ($S_2$ Stokes parameter)~\cite{kuriakose2022few}. Furthermore, control beams with opposite circular polarisation lead to $\phi$ with opposite sign (see Fig.~\ref{fig3}b). The phase shift is then given by
\begin{equation}
\phi \approx \frac{1}{2} \left( \frac{I_{D}^{\sigma^+} - I_{A}^{\sigma^+}}{I_{D}^{\sigma^+} + I_{A}^{\sigma^+}} - \frac{I_{D}^{\sigma^-} - I_{A}^{\sigma^-}}{I_{D}^{\sigma^-} + I_{A}^{\sigma^-}} \right)
\end{equation}
Here $I_{D,A}^{\sigma^+,\sigma^-}$ are sums of the curves in Fig.~\ref{fig2}(a) over a small range of times near the peak of each pulse. Further details of the data processing and phase shift extraction are given in Supplementary Discussion 4.

\subsection{Model for phase shift fitting}\label{methods:model_params}

We use the model in Supplementary discussion 5 to fit the detuning dependence of the phase shift with the following equation

\begin{equation}
\frac{\partial\phi}{\partial N\upsub{pol}}=\frac{\Re\left[\left(\delta_{p}-\frac{E\upsub{BF}}{2}-i\frac{\gamma}{2}\right)\left(S_{2,c}U_{2} + iS_{3,c}\left(2U_{1}-U_{2}\right)\right)\right]}{\left|\delta_{p}-\frac{E\upsub{BF}}{2}-i\frac{\gamma}{2}\right|^{2}}
\end{equation}

Here $N\upsub{pol}$ is the number of control polaritos, $\delta_p$ is the probe detuning relative to the average of the two birefringent split polaritons states, $E\upsub{BF}$ is the birefringent splitting, $\gamma$ is the polariton linewidth, $S_{2,c}$ and $S_{3,c}$ are the transmitted control beam Stokes parameters, $U_1$ and $U_2$ come from the frequency dependent interaction constants in Eqns.~\eqref{eq:alpha1} and \eqref{eq:alpha2} as described in the main text and $\Re$ denotes taking the real part. 

The biexciton parameters $\gamma\upsub{BX}$ and $g\upsub{BX}$ are the fitting parameters while the other parameters in the model are fixed at the values given as follows. We take $g_{\mathrm{XX}}^{\parallel}=$ 10\nlu{} and $g_{\mathrm{XX}}^{\perp}=0$ as representative values from the literature, although their contribution is small and they make little difference to the fit. $A\upsub{eff}=7.1$\um{}$^2$ was obtained from FDTD simulation (see Supplementary Discussion 1). The frequency dependent polariton linewidth $\gamma$ is shown in Fig.~\ref{fig1}c. The birefringent splitting was found to be $0.858\gamma$ (Supplementary Discussion 2B). Both lasers were tuned to the center of the birefringent split states by maximising the transmission of the control beam with circular incident polarisation. For the biexciton binding energy we took the value 2.13 meV from Ref.~\cite{takemura2017spin}. A similar value was also used in Ref.~\cite{scarpelli2024probing} while a value of $\sim 1.88$ meV was found using multidimensional coherent spectroscopy~\cite{Wilmer2015}.

\subsection{Effective interaction constant for comparison across literature}\label{methods:geff}
Previous works on polariton open access microcavities have considered only linearly polarised states, making use of a larger birefringent splitting that in our system. In this case the relevant polariton interaction energy is given by~\cite{Bleu2021} $U = \left(U_1 + U_2\right)/2$. Using our best fit values of $g\upsub{BX}$ and $\gamma\upsub{BX}$ and Eqns.~\eqref{eq:alpha1} and~\eqref{eq:alpha2} to calculate $\left|U\right|$ we obtain 4.7\uev{} at $\Delta=-0.03$ meV.

To better compare between different works we can scale out the dependences on effective mode area $A\upsub{eff}$ and frequency dependent excitonic fraction from $U$. This leaves an effective excitonic interaction constant $g\upsub{eff}\left(\Delta\right) = \left|U\right|A\upsub{eff}/\left|X\right|^4$. This incorporates all enhancements due to e.g. biexciton coupling and may therefore still be highly frequency dependent even though the $\left|X\right|^4$ dependence has been scaled out. It also incorporates both the real (blueshift) the imaginary (line broadening) parts of $\alpha_2$, both of which can contribute to polariton correlations~\cite{schnuriger2026quantum} or phase shift (see supplementary discussion 5). With the above quoted value of $U$ we obtain the value $g\upsub{eff} = 134$\nlu{} at $\Delta=-0.03$ meV quoted in Table~\ref{tab:comparison}.


\section*{Data availability}
The data supporting the findings of this study are freely available in the University of Sheffield repository with the identifier 10.15131/shef.data.31819660

\bibliography{main_bib}

\section*{Acknowledgments}
This work was supported by grant EP/V026496/1 of EPSRC of the UK. This work was partly supportedby the Paris Ile de France Région in the framework of DIM SIRTEQ, by the European Research Council (ERC) under the European Union’s Horizon 2020 research and innovation programme (project ARQADIA, grant agreement no. 949730), and under Horizon Europe research and innovation programme (ANAPOLIS, grant agreement no. 101054448). For the purpose of open access, the author has applied a Creative Commons Attribution (CC BY) licence to any Author Accepted Manuscript version arising.

\section*{Author Contributions}
P.M.W and D.N.K. designed the experiment with contributions from F.B. and A.E. P.M.W., F.B. and A.E. built the experimental apparatus and A.E. and F.B. collected and analysed the data with contributions from P.M.W. P.M.W and O.K. developed the theoretical model. S.R., J.B., A.L., and M.M. designed, fabricated and characterised the semiconductor sample. The dielectric mirror sample was designed and fabricated by J.S. T.I., F.B. and A.E. performed additional characterisation of the samples. The manuscript was written by P.M.W, F.B, A.E. and O.K. All authors contributed to the discussion and interpretation of the data and revision of the manuscript. 

\section*{Conflict of interest statement}
All authors declare that there are no conflicts of interests.



\end{document}